# Low-to-mid Al content ($x$~0-0.56) $Al_xIn_{1-x}N$ layers deposited on Si(100) by RF sputtering

*Rodrigo Blasco\*, Sirona Valdueza-Felip, Daniel Montero, Michael Sun, Javier Olea and Fernando B. Naranjo*

R. Blasco, M. Sun, Dr. S. Valdueza-Felip and Dr. F. B. Naranjo
Photonics Engineering Group, University of Alcalá, 28871 Alcalá de Henares, Spain
E-mail: rodrigo.blasco@uah.es

D. Montero and Dr. J. Olea
Applied Physics Department III, University Complutense of Madrid, 28040 Madrid, Spain

**Abstract**

Radio frequency sputtering is a low-cost technique for the deposition of large-area single-phase AlInN on silicon layers with application in photovoltaic devices. Here we study the effect of the Al mole fraction $x$ from 0 to 0.56 on the structural, morphological, electrical and optical properties of $n$-$Al_xIn_{1-x}N$ layers deposited at 550ºC on $p$-Si(100) by radio frequency sputtering. X-ray diffraction data show a wurtzite structure oriented along the $c$-axis in all samples, where the full width at half maximum of the rocking curve around the InN (0002) diffraction peak decreases from ~9º to ~3º when incorporating Al to the AlInN layer. The root mean square surface roughness, estimated from atomic force microscopy, evolves from 20 nm for InN to 1.5 nm for $Al_{0.56}In_{0.44}N$. Low-temperature photoluminescence spectra show a blue shift of the emission energy from 1.59 eV (779 nm) for InN to 1.82 eV (681 nm) for $Al_{0.35}In_{0.65}N$ accordingly to the Al content rise. Hall effect measurements of $Al_xIn_{1-x}N$ (0 < $x$ < 0.35) on sapphire samples grown simultaneously point a residual $n$-type carrier concentration in the $10^{21}$ cm$^{-3}$ range. The developed $n$-AlInN/$p$-Si junctions present promising material properties to explore their performance operating as solar cell devices.

# 1. Introduction

III-nitride semiconductors are interesting materials for the development of novel devices due to their direct bandgap energy tunable from the infrared (0.7 eV) for InN to the UV (6.2 eV) for AlN, their strong chemical and temperature endurance and their radiation hardness [1]. In particular, AlInN alloys are special candidates for power electronics, high-efficient emitters, optoelectronic devices and solar cells [2-8]. In that sense, radio frequency (RF) magnetron sputtering allows the deposition of large-area and single-phase AlInN material using this low-cost and low-temperature technology exportable to the industry.

Our group has already experience on the deposition of In-rich $Al_xIn_{1-x}N$ (x~0−0.39) on glass [9], sapphire [10] and directly on Si(111) [11] substrates and also with an AlN buffer layer [12] at different temperatures ranging from room temperature [9] to 550ºC [10,11]. We have achieved high-structural-quality nitride layers on sapphire with a bandgap energy of 1.76−2.03 eV (704−610 nm) and a room temperature photoluminescence emission that blue shifts from 1.59 eV (779 nm) to 1.86 eV (666 nm) for InN and $Al_{0.36}In_{0.64}N$ layers, respectively [10].

However, nowadays the Si technology is based on Si with (100) cubic crystal orientation due to their lower silicon atom density at the surface and amount of dangling bonds that generate non-desired recombination centers [13]. So, to adapt the III-nitrides to this technology we need to develop the growth of $Al_xIn_{1-x}N$ layers on Si(100) substrates.

InN and AlInN layers have been already grown by molecular beam epitaxy-based methods on Si(100) [14,15]. Particularly, Chen *et al.* studied the effect of In/Al ratios on the structural and optical properties of $Al_xIn_{1-x}N$ ($x$ = 0.20−0.43) films grown by RF metal-organic molecular beam epitaxy obtaining layers oriented along the *c*-axis. However, the crystalline quality of the AlInN layers degrades with increasing the Al content for $x$ > 0.36 due to the presence of structural defects [15].

Published papers on III-nitride layers deposited by RF sputtering reveal that they were mainly deposited using a mixture of nitrogen and argon, contrary as presented in this work. He *et al.* studied the material properties of InN and $Al_xIn_{1-x}N$ ($x < 0.30$) deposited on Si(100) by RF magnetron sputtering, obtaining also *c*-axis oriented layers without phase separation in the alloy. Both InN and AlInN films showed a morphology with pyramids like hillocks [16].

Afzal *et al.* reported the deposition of AlInN thin films on Si(100) through the annealing of Al and InN stacking layers deposited by RF magnetron sputtering at 100ºC [17,18]. They showed that the mixing of the Al/InN stacked layer and the surface roughness becomes more prominent with the increase of the annealing temperature from 200ºC to 800ºC [19].

On the other hand, first solar cells based on $n$-$Al_xIn_{1-x}N$ on $p$-Si(100) heterojunctions deposited by RF sputtering were developed by Liu *et al.* and showed a conversion efficiency of 1.1% under 1 sun AM-1.5G illumination for with an Al mole fraction of $x = 0.27$ and a bandgap energy of 2.1 eV [8]. However, the successful fabrication of efficient AlInN/Si devices requires the increase of the Al content of the alloy to improve the overlapping of the spectral response of the device and the maximum of the solar spectrum and increase therefore the conversion efficiency. In this sense, recent theoretical simulations on the photovoltaic performance of these kind of heterojunctions predict their high potential as solar cells pointing to a conversion efficiency of 23.6% under 1-sun AM1.5G illumination in junctions with mid-Al mole fractions ~0.49 and an antireflection coating [20].

Here we study the effect of the Al mole fraction $x$ from 0 to 0.56 on the structural, morphological, electrical and optical properties of $n$-$Al_xIn_{1-x}N$ on $p$-Si(100) layers deposited at 550ºC by RF magnetron sputtering with the aim of fabricating an operative low-cost solar cell.

## 2. Experimental Section

$Al_xIn_{1-x}N$ samples were deposited on $p$-doped 300 μm thick Si(100) with a resistivity of 1–10 Ω·cm using a reactive RF magnetron sputtering system with 2" confocal magnetron cathodes of pure In (4N5) and pure Al (5N), and pure nitrogen (6N) as reactive gas. The base pressure of the sputtering system is in the order of $10^{-7}$ to $10^{-8}$ mbar. Substrates were chemically cleaned in organic solvents before being loaded in the chamber where they were outgassed for 30 min at 600ºC. Afterwards, substrates were cooled down to the growth temperature. Prior deposition, targets and substrate were cleaned using a plasma etching with Ar (99.999%) in the growth chamber. Optimized AlInN layers were deposited with a nitrogen flow of 14 sccm at a pressure of 0.47 Pa. Reactive sputtering with a pure nitrogen atmosphere was used during the deposition process in order to optimize the crystalline quality of the layers [21]. The RF power applied to Al target, $P_{Al}$, was varied from 0 to 225 W (samples S1 to S7 in Table I), whereas the RF power applied to In target and the temperature were fixed to 30 W and 550ºC, respectively. The AlInN films are nominally ~90 nm thick. However, the real thickness of the layers was estimated from X-ray reflection (XRR) measurements pointing to an average of 88±8 nm. The thickness values obtained for each sample are listed in Table I, which shows also the linear dependence of the deposition rate of the samples vs the power applied to the Al target. Under these deposition conditions the deposition rate evolves from 0.59 nm/min for $P_{Al} = 0$ W to 1.66 nm/min for $P_{Al} = 225$ W.

The alloy mole fraction, crystalline orientation, and mosaicity of the films was evaluated by high-resolution X-ray diffraction (HRXRD) measurements using a PANalytical X'Pert Pro MRD system. Atomic force microscopy (AFM) was used to evaluate the surface morphology of the layers and estimate its root mean square (rms) surface roughness using a Bruker multimode Nanoscope III A microscope in tapping mode. The electrical properties were analysed in AlInN layers simultaneously deposited on sapphire substrates under the same growth conditions. We used room-temperature Hall-effect measurements under the conventional van der Pauw geometry. Temperature dependent (11−300 K) photoluminescence (PL) measurements were carried out by exciting the samples with 30 mW of a continuous-wave diode laser emitting at λ = 405 nm focused onto a 1-mm diameter spot. The emission was collected with a 193-mm-focal-length Andor spectrograph equipped with a UV-extended silicon-based charge-coupled-device camera operating at -60ºC.

## 3. Results and Discussion

The structural quality of the AlInN films was evaluated by HRXRD measurements. All layers present a wurtzite crystalline structure highly oriented along the *c*-axis perpendicular to the sample surface, as shown in the 2θ/ω diffractograms of Fig. 1(a). This figure displays the Si(100) and the AlInN (0002) diffraction peaks showing that there is no phase separation in the AlInN material along the studied range of $P_{Al}$. The increase of $P_{Al}$ shifts the (0002) reflection peak towards high diffraction angles, which means that there is a reduction of the c lattice parameter from 5.729Å for InN (S1) to 5.299 Å for AlInN deposited with $P_{Al}$ = 225 W (S7). Vegard's law [22] was applied to determine the average Al composition of the ternary alloy via estimation of c-lattice parameter from HRXRD. If we assume that the $Al_xIn_{1-x}N$ layers are fully relaxed, we can deduced their Al mole fraction *x* which linearly increases with $P_{Al}$ in the range of *x*~0–0.56, as displayed in Fig. 1(b) and Table I. Moreover, the crystalline quality of the AlInN layers improves with the Al incorporation, being attributed to the increase of the adatom mobility with the Al power supply. Moreover, the full width at half maximum of the rocking curve around the AlInN (0002) diffraction peak decreases from ~9º for the InN layer to ~3º for AlInN layers even for low Al RF powers, as shown in Fig. 1(c). These values are on the same order of magnitude of the ones obtained by Núñez-Cascajero *et al.* in similar AlInN on Si(111) samples [11].

This reduction in the AlInN mosaicity with the Al is accompanied with a change in the surface morphology of the samples, which rms surface roughness estimated from AFM images evolves from 20 nm for InN (S1) to 2.0 nm and 1.5 nm for $Al_{0.35}In_{0.65}N$ (S4) and $Al_{0.56}In_{0.44}N$ layers (S7), respectively. Figures 2 (a), (b) and (c) show the 2×2 μm² AFM pictographs of these three samples. Besides, as observed in previous results of AlInN alloys deposited by RF sputtering [10,11], this reduction in the AlInN mosaicity and in the surface roughness with the Al, may entail a change in the surface morphology of the samples, from closely-columnar for InN to compact for $Al_xIn_{1-x}N$."

A proper control of *n*-type doping is of crucial importance to develop efficient devices. Hall effect measurements were carried out in $Al_xIn_{1-x}N$ films simultaneously deposited on sapphire to explore the evolution of their electrical characteristics, such as carrier concentration, layer resistivity and carrier mobility as a function of the alloy mole fraction. As summarized in Table I, we estimated an increase of the resistivity from 0.2 mΩ·cm to 3.4 Ω·cm for the InN (S1) to the $Al_{0.45}In_{0.55}N$ samples (S5), respectively. Moreover, the residual *n*-type carrier concentration drops two orders of magnitude from $7.5 \times 10^{21}$ cm$^{-3}$ to $1.6 \times 10^{19}$ cm$^{-3}$, and the carrier mobility drops from 4.2 cm$^2$/V·s to 0.1 cm$^2$/V·s within the same Al mole fraction range. For $x > 0.45$ electrical measurements were not reliable due to the high resistivity of the layers, above the resolution of the Hall effect setup. The high carrier concentration presented in these (Al)InN layers is related to an unintentional doping coming from hydrogen and oxygen impurities during the sputtering deposition and point defects like nitrogen vacancies as is observed by [23, 24, 25].

Low-temperature (T = 11 K) PL measurements of AlInN samples grown on Si(100) reveal a dominant PL emission coming from the $Al_xIn_{1-x}N$ layers with Al mole fractions below 0.35, as plotted in Fig. 3. No PL emission was observed coming from AlInN layers with $x > 0.35$. The high layer resistivity, low *n*-type carrier concentration and poor carrier mobility present in $Al_xIn_{1-x}N$ on Si structures with $x > 0.35$ are probably the reasons why there is not PL emission coming from the material. The PL emission energy blue shifts with the Al content from 1.59 eV (779 nm) for the InN (S1) to 1.82 eV (681 nm) for the $Al_{0.35}In_{0.65}N$ layer (S4), showing a similar broadening of the emission with a FWHM of ~300 meV in all cases. In order to investigate the origin of the luminescence, the thermal evolution of the PL peak energy was studied between 11 K and room temperature for InN (S1) and Al$_{0.28}$In$_{0.72}$N samples (S3).

Figures 4(a) and 5(a) show the evolution of the PL emission for the InN and the Al$_{0.28}$In$_{0.72}$N on Si(100) samples, respectively. The latter was chosen since it showed the highest optical quality in terms of emission intensity and FWHM.

Concerning the InN sample [see Fig. 4(b)], an almost constant emission energy is measured independently of the temperature for the analyzed range, with a weak S-shape evolution and a red-shift or the emission as low as ~5 meV between 11 K and room temperature. On the contrary, the PL emission of the Al$_{0.28}$In$_{0.72}$N sample [see Fig. 5(b)] shows a clear red shift of the emission energy of ~35 meV between 11 K and room temperature. At the same time, the FWHM of the emission peak at room temperature are 390 meV and 420 meV for the InN (S1) and Al$_{0.28}$In$_{0.72}$N samples (S3), respectively. This points out an emission origin related with an energy-band of impurities, in agreement with the high carrier concentration measured in the samples. This behavior has been previously observed in both InN and AlInN samples deposited by RF sputtering on sapphire [10] being attributed to the existence of an important concentration of impurities like oxygen and probably nitrogen vacancies in the layers.

It is remarkable the reduced effect of the alloy disorder in the FWHM of the observed PL when increasing the Al content in the layers. This effect is attributed to the high growth temperature used in this study (550ºC) that promotes the adatom diffusion, as reflected in the aforementioned reduction of the rms surface roughness measured by AFM.

Figure 6 shows the thermal quenching of the normalized integrated PL intensity for the InN and Al$_{0.28}$In$_{0.72}$N samples. The activation energy of the main process involved in this emission quenching, E$_A$, was estimated using equation $I(T) = I_0/\left(1 + ae^{\frac{-E_A}{k_B T}}\right)$, where I$_0$ is the integrated intensity at T = 0 K, k$_B$T is the thermal energy, and *a* is a fitting constant that is related to the nonradiative-to-radiative recombination ratio of the process. Activation energies of $E_A = 20 \pm 2.0$ meV and $E_A = 24 \pm 4.4$ meV were estimated for the InN and $Al_{0.28}In_{0.72}N$ samples, respectively. The value obtained for InN is lower than the one of $E_A = 51$ meV estimated in InN on sapphire samples deposited at lower substrate temperature (300ºC) [26]. This difference can be attributed to a change in the formation of impurity complexes with the temperature. However, the obtained value for the $Al_{0.28}In_{0.72}N$ is in agreement with the one estimated in AlInN on Si(111) samples with similar Al mole fraction [11]. The extracted fitting parameters are $a = 0.6$ and $a = 1.9$ for InN and $Al_{0.28}In_{0.72}N$ samples, respectively.

## 4. Conclusion

In this work we demonstrated the feasibility of depositing high-quality and single-phase *n*-type $Al_xIn_{1-x}N$ layers with an Al mole fraction ranging from x = 0 to 0.56 on *p*-Si(100) substrates by RF magnetron sputtering. The increase of the Al mole fraction results in an improvement of the structural and morphological quality of the layers, showing a minimum FWHM of the (0002) AlInN rocking curve of ~3º and a minimum rms surface roughness of 1.5 nm for $x = 0.56$. For Al mole fractions up to $x = 0.35$, samples show strong low- and room-temperature photoluminescence emission, which blue shifts from 1.59 eV (779 nm) for InN to 1.82 eV (681 nm) for $Al_{0.35}In_{0.65}N$ samples at 11 K. Within this Al mole fraction range ($0 < x < 0.35$), the $Al_xIn_{1-x}N$ resistivity increases from 0.2 to 4.6 mΩ·cm, and the carrier concentration and mobility drop from $7.5 \times 10^{21}$ to $1.0 \times 10^{21}$ cm$^{-3}$ and from 4.2 to 1.4 cm$^2$/V·s, respectively. The developed *n*-AlInN on *p*-Si(100) junctions deposited by RF sputtering present promising material properties to explore their performance operating as solar cell devices up to 35% of Al.

## Acknowledgements

Financial support was provided by national projects ANOMALOS (TEC2015-71127-C2-2-R), TEC2017-84378-R and NERA (RTI2018-101037-B-I00); projects from the Comunidad de

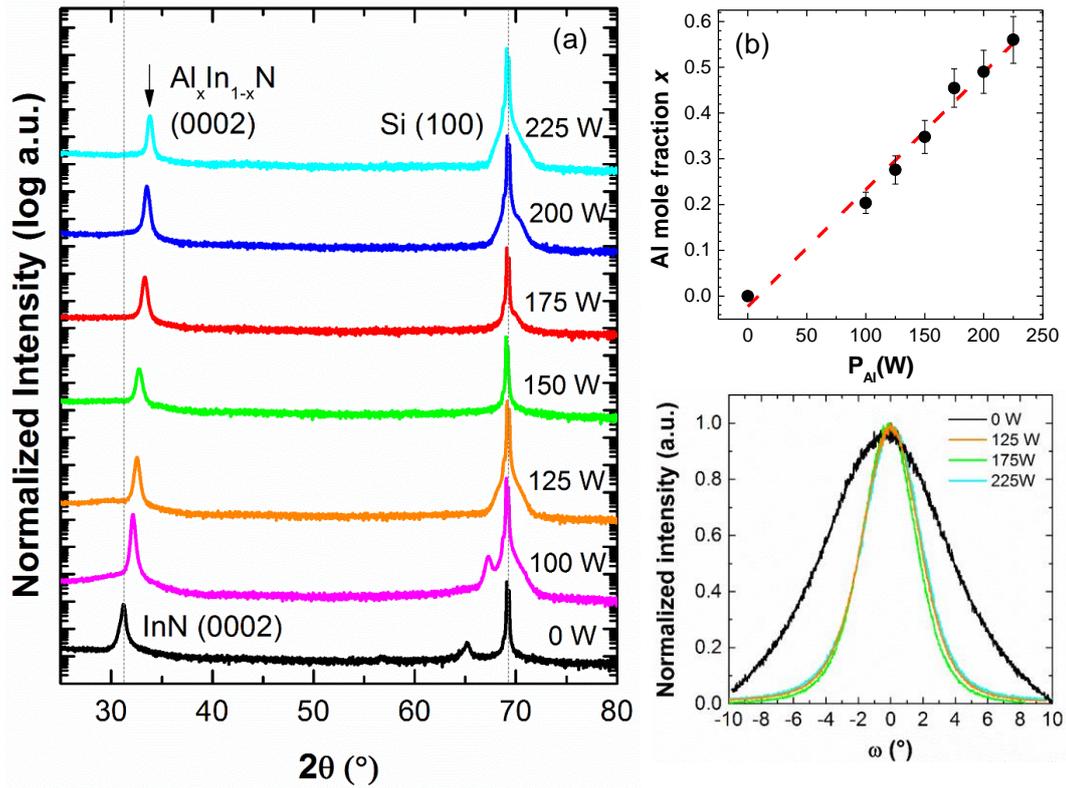

**Figure 1.** (a) 2θ/ω scans of the AlInN on Si(100) structures vs $P_{Al}$. (b) Al mole fraction *x* of the $Al_xIn_{1-x}N$ layers estimated from HRXRD measurements as a function of $P_{Al}$. The Al mole fraction estimation owns an error bar of ~1–4% depending on the Al mole fraction, as presented in Ref. [10]. (c) Normalized intensity of the rocking curve of the AlInN (0002) diffraction peak vs $P_{Al}$.

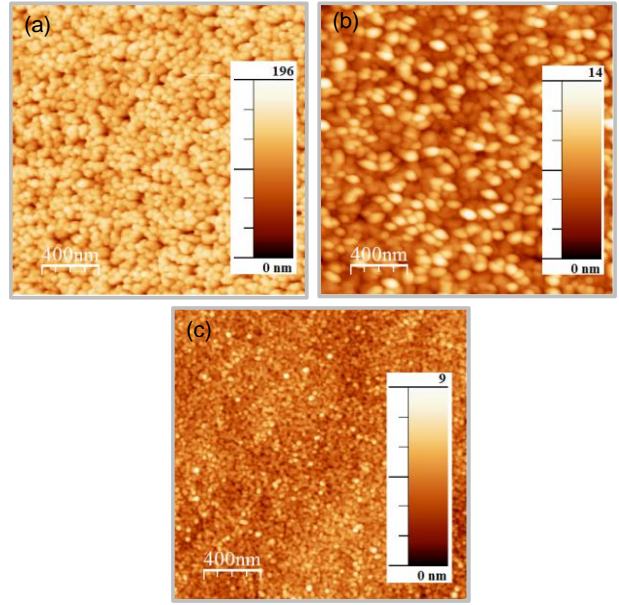

**Figure 2.** AFM pictographs of 2×2 μm² area of the (a) InN (S1), (b) Al$_{0.35}$In$_{0.65}$N (S4) and (c) Al$_{0.56}$In$_{0.44}$N on Si(100) samples (S7). Rms surface roughness values are 20, 2.0 and 1.5 nm, respectively.

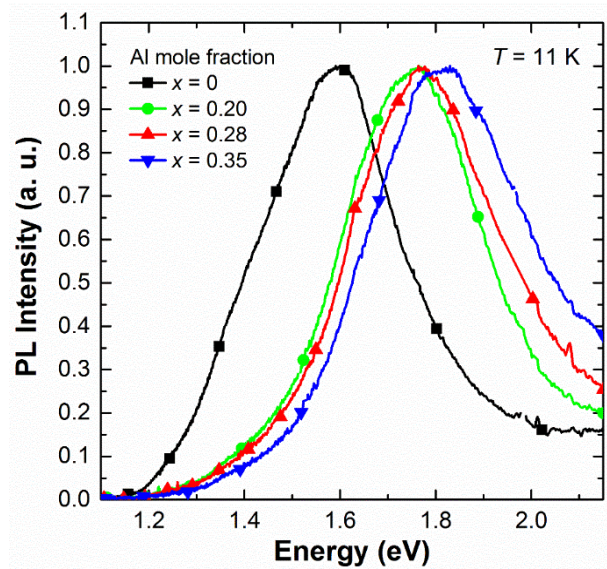

**Figure 3.** Normalized low-temperature PL emission of the Al$_x$In$_{1-x}$N on Si(100) samples vs the Al mole fraction. For $x > 0.35$ no PL emission was observed.

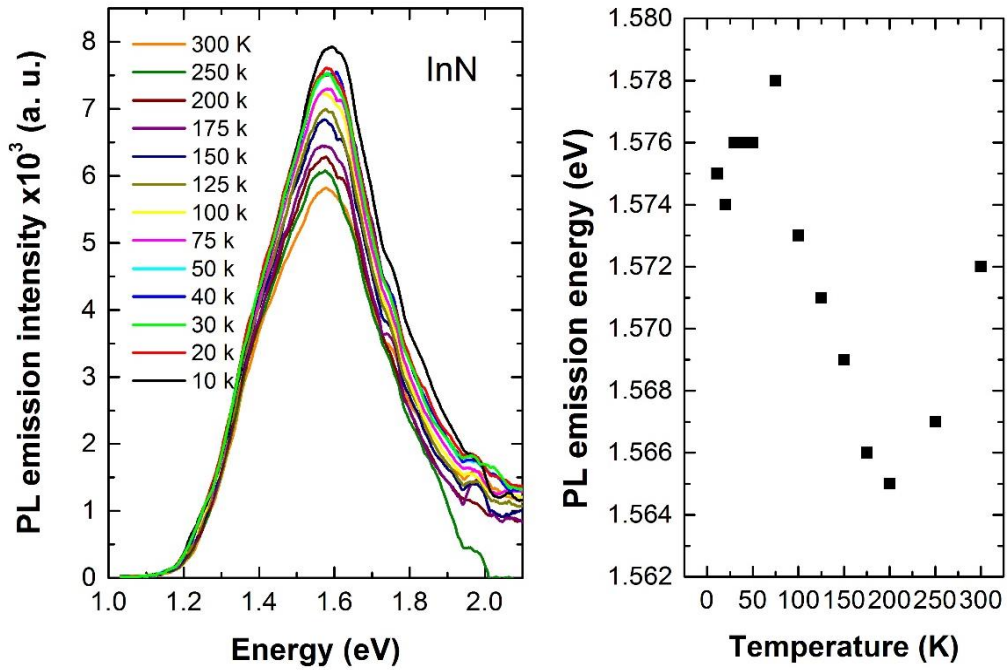

**Figure 4.** (Left) PL emission spectra as a function of the temperature for the InN on Si(100) sample S1. (Right) Evolution of the PL emission energy vs the temperature of the same sample.

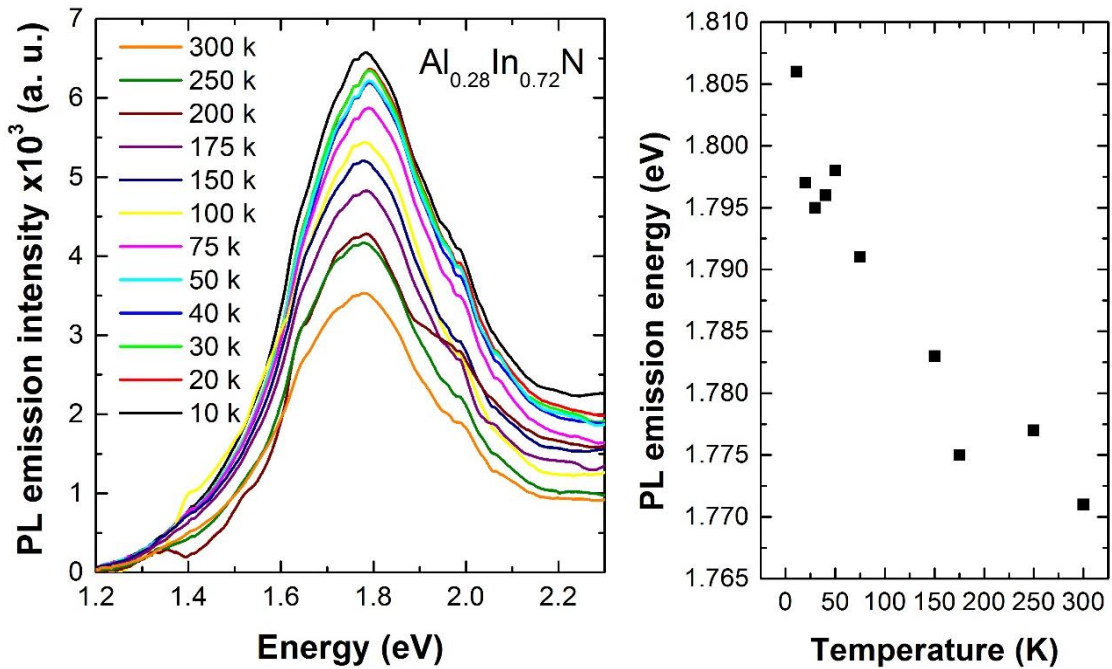

**Figure 5.** (Left) PL emission spectra as a function of the temperature for the $Al_{0.28}In_{0.72}N$ on Si(100) sample S3. (Right) Evolution of the PL emission energy vs the temperature of the same sample.

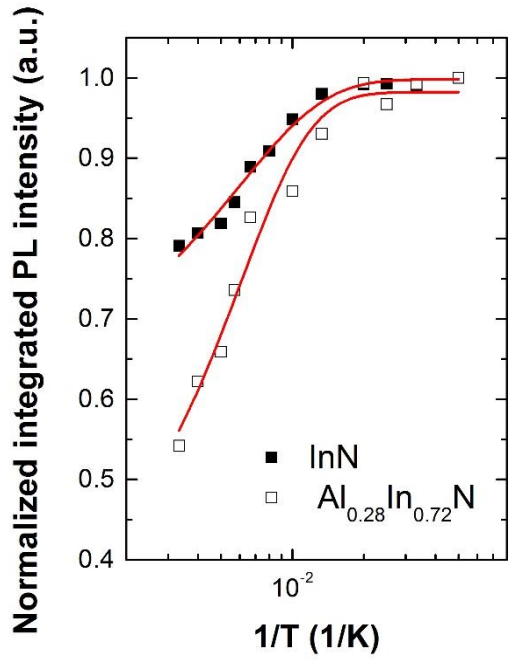

**Figure 6.** Evolution of the normalized integrated PL intensity as a function of the temperature for the InN (S1) and the Al$_{0.28}$In$_{0.72}$N on Si(100) samples (S3).

**Table 1.** Summary of the structural, morphological, optical and electrical characterization results of the Al$_x$In$_{1-x}$N samples: RF power applied to the Al target (P$_{Al}$); thickness estimated by XRR; deposition rate; Al mole fraction $x$ and c lattice parameter estimated from HRXRD; low-temperature PL emission energy (E$_{PL}$); room-temperature carrier concentration (n), resistivity (ρ) and mobility (μ) estimated via Hall effect.

| Sample | P$_{Al}$ (W) | Thickness (nm) | Deposition rate (nm/min) | $x$ | c (Å) | E$_{PL}$ (eV) | n (cm$^{-3}$) | ρ (Ω·cm) | μ (cm$^2$/V·s) |
|---|---|---|---|---|---|---|---|---|---|
| S1 | 0 | 88 | 0.59 | 0 | 5.729 | 1.59 | 7.5×10$^{21}$ | 2.0×10$^{-4}$ | 4.2 |
| S2 | 100 | 91 | 0.77 | 0.20 | 5.556 | 1.75 | 2.2×10$^{21}$ | 9.0×10$^{-4}$ | 3.2 |
| S3 | 125 | 83 | 0.83 | 0.28 | 5.504 | 1.77 | 1.2×10$^{21}$ | 2.5×10$^{-3}$ | 2.1 |
| S4 | 150 | 95 | 1.03 | 0.35 | 5.452 | 1.82 | 1.0×10$^{21}$ | 4.6×10$^{-3}$ | 1.4 |
| S5 | 175 | 80 | 1.16 | 0.45 | 5.375 | - | 1.6×10$^{19}$ | 3.4 | 0.1 |
| S6 | 200 | 87 | 1.36 | 0.49 | 5.350 | - | - | - | - |
| S7 | 225 | 88 | 1.66 | 0.56 | 5.299 | - | - | - | - |